\documentclass[twocolumn,prb,color,superscriptaddress,psfig]{revtex4}
\usepackage{graphicx}
\usepackage{epsfig}
\begin{document}
\title{Dzyaloshinsky-Moriya antisymmetric exchange coupling in cuprates: Oxygen effects}
\author{A.S.\ Moskvin}
\affiliation{Ural State University, 620083 Ekaterinburg,  
Russia}
\date{\today}
\begin{abstract}
 We revisit a problem of  Dzyaloshinsky-Moriya antisymmetric exchange coupling for a single bond in cuprates specifying the local spin-orbital contributions to Dzyaloshinsky vector focusing on the oxygen term. The Dzyaloshinsky vector and respective weak ferromagnetic moment is shown to be a superposition of comparable and, sometimes, competing local Cu and O contributions. The intermediate oxygen $^{17}$O Knight shift is shown to be an effective tool to inspect the effects of Dzyaloshinsky-Moriya coupling in an external magnetic field. We predict the effect of $strong$ oxygen weak antiferromagnetism in edge-shared CuO$_2$ chains due to uncompensated oxygen Dzyaloshinsky vectors. Finally, we revisit the effects of symmetric spin anisotropy, in particular, those directly induced by Dzyaloshinsky-Moriya coupling.
\end{abstract}

\maketitle

\section{Introduction}

Fifthy years ago Borovik-Romanov and Orlova \cite{WF} have discovered the phenomenon of weak ferromagnetism which origin was shortly after \cite{Dzyaloshinsky} bound up with exchange-relativistic effect mainly with antisymmetric exchange coupling.
Starting from pioneer papers by Dzyaloshinsky \cite{Dzyaloshinsky} and Moriya \cite{Moriya} the 
Dzyaloshinsky-Moriya (DM) antisymmetric exchange coupling was extensively investigated in 60-80ths in connection with weak ferromagnetism focusing on  hematite $\alpha$-Fe$_2$O$_3$ and orthoferrites RFeO$_3$.\cite{Moskvin} The stimulus to a renewed interest to the subject was given by the cuprate problem, in particular, by the weak ferromagnetism observed in La$_2$CuO$_4$\cite{Thio} and  many other interesting effects for the DM systems, in particular, the "field-induced gap" phenomena.\cite{Affleck}  At variance with typical 3D systems such as orthoferrites, cuprates  are characterised by a low-dimensionality, large diversity of Cu-O-Cu bonds including corner- and edge-sharing, different ladder configurations, strong quantum effects for $s=1/2$ Cu$^{2+}$ centers, and a particularly strong Cu-O covalency resulting in a comparable magnitude of hole charge/spin densities on copper and oxygen sites. 
Several groups (see, e.g., Refs.\onlinecite{Coffey,Koshibae,Shekhtman}) developed the microscopic model approach by Moriya for different 1D and 2D cuprates, making use of different perturbation schemes, different types of low-symmetry crystalline field, different approaches to intra-atomic electron-electron repulsion. However, despite a rather large number of publications and hot debates (see, e.g., Ref.\onlinecite{debate}) the problem of exchange-relativistic effects, that is of antisymmetric exchange and related problem of spin anisotropy in cuprates remains to be open (see, e.g., Refs.\onlinecite{Tsukada,Kataev} for recent experimental data and discussion). Common shortcomings of current approaches to DM coupling in 3d oxides concern a problem of allocation of Dzyaloshinsky vector and respective "weak" (anti)ferromagnetic moments, and  full neglect of spin-orbital effects for "nonmagnetic" oxygen O$^{2-}$ ions, which are usually believed to play only  indirect intervening role.
From the other hand, the oxygen $^{17}$O NMR-NQR studies of weak ferromagnet La$_2$CuO$_4$\cite{Walstedt}  seem  to evidence unconventional local oxygen "weak-ferromagnetic" polarization which origin cannot be explained in frames of current models.
 All this stimulated the critical revisit of many old  approaches to the spin-orbital effects in 3d oxides, starting from the choice of proper perturbation scheme and effective spin Hamiltonian model, implied usually  only  indirect intervening role played by "nonmagnetic" oxygen O$^{2-}$ ions. 
 
 In this paper we revisit a problem of  DM antisymmetric exchange coupling for a single bond in cuprates specifying the local spin-orbital contributions to Dzyaloshinsky vector focusing on the oxygen term. In Sec.II we present a short overview of the spin-Hamiltonian of a typical three-center (Cu-O-Cu) two-hole system. Microscopic theory of DM coupling is presented in Sec.III.  The Dzyaloshinsky vector is shown to be a superposition of comparable and, sometimes, competing local Cu and O contributions. In Sec.IV we examine a response of DM coupled Cu$_1$-O-Cu$_2$ bond on an uniform and staggered external fields, and demonstrate some unusual manifestations of local oxygen contribution to DM coupling. The intermediate oxygen $^{17}$O Knight shift is shown to be an effective tool to inspect the effects of DM coupling in an external magnetic field.  In Sec.V we revisit a problem of symmetric spin anisotropy with inclusion of local oxygen spin-orbital contributions. 
\begin{figure}[b]
\includegraphics[width=8.5cm,angle=0]{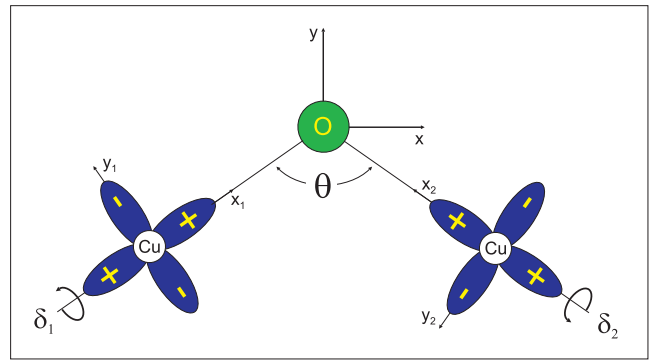}
\caption{Geometry of the three-center (Cu-O-Cu) two-hole system with ground Cu 3d$_{x^2-y^2}$ states.}
\label{fig1}
\end{figure}  
 
 \section{Spin-Hamiltonian}

 For illustration, below we address a typical for cuprates the three-center (Cu$_1$-O-Cu$_2$) two-hole system with tetragonal Cu on-site symmetry and ground Cu 3d$_{x^2-y^2}$ states (see Fig. 1) which conventional bilinear spin Hamiltonian is written in terms of copper spins as follows
\begin{equation}
\hat H_s(12)=J_{12}(\hat{\bf s}_1\cdot \hat{\bf s}_2)+{\bf D}_{12}\cdot [\hat{\bf s}_1\times \hat{\bf s}_2]+\hat{\bf s}_1{\bf \stackrel{\leftrightarrow}{K}}_{12} \,\hat{\bf s}_2 \, ,\label{1}
\end{equation} 
 where $J_{12}>0$ is an exchange integral, ${\bf D}_{12}$ is the Dzyaloshinsky vector, ${\bf \stackrel{\leftrightarrow}{K}}_{12}$ is a symmetric second-rank tensor of the anisotropy constants. 
 In contrast with $J_{12}, {\bf \stackrel{\leftrightarrow}{K}}_{12}$, the Dzyaloshinsky vector ${\bf D}_{12}$
 is antisymmetric with regard to the site permutation: ${\bf D}_{12}=-{\bf D}_{21}$. Hereafter we will denote $J_{12}=J,{\bf \stackrel{\leftrightarrow}{K}}_{12}={\bf \stackrel{\leftrightarrow}{K}},{\bf D}_{12}={\bf D}$, respectively.
 It should be noted that making use of effective spin Hamiltonian (\ref{1}) implies a removal of orbital degree of freedom that calls for a caution with DM coupling as it changes both a spin multiplicity,  and an orbital state.

  It is clear that the applicability of such an operator as $\hat H_s(12)$ to describe all the "oxygen" effects is extremely limited. Moreover, the question arises in what concerns the composite structure and spatial distribution of what that be termed as the Dzyaloshinsky vector density. Usually this vector is assumed to be located on the bond connecting spins 1 and 2. 
  
  Strictly speaking, spin Hamiltonian $H_s(12)$ can be viewed as a result of the projection onto the purely ionic Cu$_1^{2+}(3d_{x^2-y^2})$-O$^{2-}(2p^6)$-Cu$_2^{2+}(3d_{x^2-y^2})$ ground state of the two-hole spin Hamiltonian
 \begin{widetext}
\begin{equation}
\hat	H_s=\sum_{i<j}I(i,j)(\hat{\bf s}(i)\cdot \hat{\bf s}(j))+\sum_{i<j}({\bf d}(i,j)\cdot [\hat{\bf s}(i)\times \hat{\bf s}(j)])+\sum_{i<j}\hat{\bf s}(i){\bf \stackrel{\leftrightarrow}{K}}(i,j)\, \hat{\bf s}(j) \, ,
\end{equation}
 \end{widetext}
 which  implies not only both copper and oxygen hole location, but allows to account for purely oxygen two-hole configurations. Moreover, such a form allows us to neatly separate both one-center and two-center effects. 
 Two-hole spin Hamiltonian can be projected onto three-center states incorporating the Cu-O charge transfer effects. 
 
 For a composite two $s=1/2$ spins system one should consider three types of the vector order parameters:
\begin{equation}
\hat{\bf S}=\hat{\bf s}_1+\hat{\bf s}_2;\,\hat{\bf V}=\hat{\bf s}_1-\hat{\bf s}_2;\,\hat{\bf T}=2[\hat{\bf s}_1\times \hat{\bf s}_2]
\end{equation}
with a kinematic constraint:
 \begin{equation}
 \hat{\bf S}^2+\hat{\bf V}^2=3\hat{\bf I};\,(\hat{\bf S}\cdot \hat{\bf V})=0;\,
(\hat{\bf T}\cdot \hat{\bf V})=6i; \, [\hat{\bf T}\times \hat{\bf V}]=\hat{\bf S}.
\label{SVT}
\end{equation}
In a sense the $\hat{\bf V}$ operator describes  the effect of local antiferromagnetic order, while $\hat{\bf T}$ operator may be associated with a vector chirality.\cite{Kolezhuk} In recent years, phases with broken vector chirality in frustrated quantum spin chains have attracted considerable
interest. Such phases are characterized by nonzero long-range correlations of the vector order parameter $\langle \hat{\bf T}\rangle $. Interestingly that  a chirally ordered phase can manifest itself as a "nonmagnetic" one, with 
$\langle \hat{\bf S}\rangle =\langle \hat{\bf V}\rangle =0$.

Both $\hat{\bf T}$ and $\hat{\bf V}$ operators  change the spin multiplicity with matrix elements  
$$
\langle 00|\hat T_m|1n\rangle =-\langle 1n|\hat T_m|00\rangle =i\delta_{mn};
$$
 \begin{equation}
\langle 00|\hat V_m|1n\rangle =\langle 1n|\hat V_m|00\rangle =\delta_{mn},
\end{equation}
 where we made use of Cartesian basis for $S=1$. The eigenstates of the operators  $\hat{V}_n$ and $\hat{T}_n$ with nonzero eigenvalues $\pm 1$ form N\'{e}el doublets $\frac{1}{\sqrt{2}}(|00\rangle \pm|1n\rangle )$ and DM doublets $\frac{1}{\sqrt{2}}(|00\rangle \pm i|1n\rangle )$, respectively. The N\'{e}el doublets correspond to classical collinear antiferromagnetic spin configurations, while the DM doublets correspond to quantum spin configurations which sometimes are associated with a rectangular 90$^0$ spin ordering in the plane orthogonal to the Dzyaloshinsky vector.

It should be noted that the both above spin Hamiltonians can be reduced to within a constant to a spin operator acting in a total spin space
 \begin{equation}
\hat	H_S=\frac{1}{4}J(\hat{\bf S}^2-\hat{\bf V}^2)+\frac{1}{2}({\bf D}\cdot \hat{\bf T})+\frac{1}{4}\hat{\bf S}{\bf \stackrel{\leftrightarrow}{K}}^S \hat{\bf S}-\frac{1}{4}\hat{\bf V}{\bf \stackrel{\leftrightarrow}{K}}^V \hat{\bf V} \, .\label{HS}
\end{equation}
 For simple dipole-like two-ion anisotropy as in Exp. (1) ${\bf \stackrel{\leftrightarrow}{K}}^S={\bf \stackrel{\leftrightarrow}{K}}^V={\bf \stackrel{\leftrightarrow}{K}}$, though in general these tensorial parameters can differ from each other. 
Making use of the anticommutator relations
\begin{equation}
\{\hat S_i,\hat S_j\}+\{\hat V_i,\hat V_j\}=2\delta_{ij};\, \{\hat V_i,\hat V_j\}=\{\hat T_i,\hat T_j\},\label{SV}	
\end{equation}
we conclude that the effective operator of symmetric anisotropy can be equivalently expressed in terms of symmetric products $\{\hat S_i,\hat S_j\}$, $\{\hat V_i,\hat V_j\}$, or $\{\hat T_i,\hat T_j\}$.

The most general form of the spin Hamiltonian (\ref{HS}) does not discriminate against copper or oxygen contribution and can be used to properly account for oxygen effects. As we will see below the ${\bf D}$ and ${\bf \stackrel{\leftrightarrow}{K}}$ parameters allow the correct separation of local copper and oxygen contributions.

It should be noted that the total spin representation and overall quantum approach is especially efficient to describe antisymmetric exchange in Cu-Cu dimer systems (see, e.g. Ref.\onlinecite{Room} and references therein).
Classical approach to $s=1/2$ spin-Hamiltonian should be applied with a caution, particularly for 1D systems.

 Before going to microscopic analysis we should note that the interaction of our three-center system with external spins and/or fields $\hat H_{ext}$ is usually addressed by introducing only two types of effective external fields: the conventional Zeeman-like uniform field and unconventional N\'{e}el-like staggered field, so that  $\hat H_{ext}$ reads as follows
 \begin{equation}
\hat	H_{ext}=-({\bf h}^S\cdot \hat {\bf S})-({\bf h}^V\cdot \hat{\bf V}).
\end{equation}
 It should be noted that an ideal N\'{e}el state is  attainable only in the limit of infinitely large staggered field, therefore for a finite staggered field ${\bf h}^V\parallel {\bf n}$ the ground state is a superposition of a spin singlet and a N\'{e}el state, 
$$
\Psi =\cos\alpha |00\rangle +\sin\alpha |1n\rangle,\, \tan2\alpha=\frac{2h^V}{J},
$$ 
 which composition reflects the role of quantum effects. For instance, in a Heisenberg spin 1/2 chain with $nn$ exchange the maximal value of staggered field $h^V=J/2$ hence the $\Psi$ function strongly differs from that of N\'{e}el state ($\langle \hat V_n\rangle =\sin2\alpha=\frac{1}{\sqrt{2}}$), and quantum mechanical average for a single spin $\langle s_z\rangle \leq\frac{1}{2}\sin\pi/4=\frac{1}{\sqrt{2}}\cdot\frac{1}{2}\approx 0.71\cdot\frac{1}{2}$ deviates strongly from classical value $\frac{1}{2}$. It should be noted that for the isolated antiferromagnetically coupled spin pair the zero-temperature uniform spin susceptibility turns into zero: $\chi ^S=0$, while for the staggered spin susceptibility we obtain $\chi ^V=2/J$.
 
 \section{Microscopic theory of DM coupling in cuprates}
 
 \subsection{Preliminaries}
 
 For the microscopic expression for Dzyaloshinsky vector to derive Moriya \cite{Moriya} made use of the Anderson's formalism of superexchange interaction \cite{PWA} with two main contributions of so called kinetic and potential exchange, respectively. Then he took into account the spin-orbital corrections to the effective d-d transfer integral and  potential exchange.  Such an approach seems to be improper to account for purely oxygen effects.
 In later papers (see, e.g. Refs.\onlinecite{Shekhtman-93,Koshibae}) the authors made use of the Moriya scheme to account for spin-orbital corrections to p-d transfer integral, however, without any analysis of oxygen contribution.
It is worth noting that in both instances the spin-orbital renormalization of a single hole transfer integral leads immediately to a lot of problems with correct responsiveness of the on-site Coulomb hole-hole correlation effects.
 Anyway the effective DM spin-Hamiltonian evolves from the high-order perturbation effects that makes its analysis rather involved and leads to many misleading conclusions.

 At variance with the Moriya approach  we start with the construction of spin-singlet and spin-triplet wave functions for our three-center two-hole system taking account of the p-d hopping, on-site hole-hole repulsion, and crystal field effects for  excited configurations $\{n\}$ (011, 110, 020, 200, 002) with different hole occupation of Cu$_1$, O, and Cu$_2$ sites, respectively. The p-d hopping for Cu-O bond implies a conventional Hamiltonian
 \begin{equation}
 \hat H_{pd}=\sum_{\alpha \beta}t_{p\alpha d\beta}{\hat p}^{\dagger}_{\alpha}{\hat d}_{\beta}+h.c.\, ,
 \label{Hpd}
\end{equation}
 where ${\hat p}^{\dagger}_{\alpha }$ creates a hole in the $\alpha $ state on the oxygen site, while  ${\hat d}_{\beta}$
annihilates a hole  in the $\beta $ state on the copper site; $t_{p\alpha d\beta}$ is a pd-transfer integral ($t_{p_x d_{x^2-y^2}}=\frac{\sqrt{3}}{2}t_{p_z d_{z^2}}=t_{pd\sigma}>0,t_{p_y d_{xy}}=t_{pd\pi}>0$).
 
 For basic 101 configuration with two $d_{x^2-y^2}$ holes localized on its parent sites we arrive at a perturbed wave function as follows
 \begin{equation}
	\Psi_{101;SM}=\Phi_{101;SM}+\sum_{\{n\}\Gamma}c_{\{n\}}({}^{2S+1}\Gamma)\Phi_{\{n\};\Gamma SM},
	\label{Psi}
\end{equation}
where the summation runs both on different configurations and different orbital $\Gamma$ states. It is worth noting that the probability amplitudes $c_{\{011\}}, c_{\{110\}}\propto t_{pd}, c_{\{200\}}, c_{\{020\}}, c_{\{002\}}\propto t_{pd}^2$.
To account for orbital effects for Cu$_{1,2}$ 3d holes and the covalency induced mixing of different orbital states for 101 configuration we should introduce an effective exchange Hamiltonian
\begin{equation}
 \hat H_{ex}=\frac{1}{2}\sum_{\alpha \beta \gamma \delta \mu \mu^{\prime}}J(\alpha \beta \gamma \delta){\hat d}^{\dagger}_{1\alpha \mu}{\hat d}^{\dagger}_{2\beta \mu^{\prime}}{\hat d}_{2\gamma \mu}{\hat d}_{1\delta \mu^{\prime}} +h.c.
 \label{ex}
 \end{equation}
Here ${\hat d}^{\dagger}_{1\alpha \mu}$ creates a hole in the $\alpha $th 3d orbital on Cu$_1$ site with spin projection $\mu$. Exchange Hamiltonian (\ref{ex}) involves both spinless and spin-dependent terms, however, it preserves the spin multiplicity of Cu$_1$-O-Cu$_2$ system. Exchange parameters $J(\alpha \beta \gamma \delta)$ are of the order of $t_{pd}^4$. 

 Then we introduce a standard effective spin Hamiltonian operating in a fourfold spin-degenerated space of basic 101 configuration with two $d_{x^2-y^2}$ holes and  can straightforwardly calculate the singlet-triplet separation to find effective exchange integral $J_{12}=J(d_{x^2-y^2}d_{x^2-y^2}d_{x^2-y^2}d_{x^2-y^2})$,  calculate the singlet-triplet mixing due to  three $local$ spin-orbital terms, $V_{so}(Cu_1),V_{so}(O),V_{so}(Cu_2)$, respectively,  to find  the $local$ contributions to Dzyaloshinsky vector: 
 \begin{equation}
 {\bf D}={\bf D}^{(1)}+{\bf D}^{(0)}+{\bf D}^{(2)}.
 \end{equation}
 Local spin-orbital coupling is taken as follows:  
$$
V_{so}=\sum_i \xi_{nl}({\bf l}_i\cdot{\bf s}_i)=\frac{\xi _{nl}}{2}[({\bf \hat l}_1+{\bf \hat l}_2)\cdot {\bf \hat S}+({\bf \hat l}_1-{\bf \hat l}_2)\cdot {\bf \hat V}]
$$
 \begin{equation}
={\bf \hat\Lambda}^S\cdot {\bf \hat S}+{\bf \hat\Lambda}^V\cdot {\bf \hat V}
\label{V_SO}
\end{equation}
 with a single particle constant $\xi_{nl}>0$ for electrons and $\xi_{nl}<0$ for holes. 
 We make use of orbital matrix elements: for Cu 3d holes $\langle d_{x^2-y^2}|l_x|d_{yz}\rangle =\langle d_{x^2-y^2}|l_y|d_{xz}\rangle =i, \langle d_{x^2-y^2}|l_z|d_{xy}\rangle =-2i$, $\langle i|l_j|k\rangle =-i\epsilon _{ijk}$ with Cu 3d$_{yz}$=$|1\rangle$, 3d$_{xz}$=$|2\rangle$ 3d$_{xy}$=$|3\rangle$, and for O 2p holes $\langle p_{i}|l_j|p_{k}\rangle =i\epsilon _{ijk}$. 
 Free cuprous Cu$^{2+}$ ion is described by a large spin-orbital coupling with $|\xi _{3d}|\cong 0.1$ eV (see, e.g., Ref.\onlinecite{Low}), though its value may be significantly reduced in oxides. Information regarding the $\xi _{2p}$ value for the oxygen O$^{2-}$ ion in oxides is scant if any. Usually one considers the spin-orbital coupling on the oxygen to be much smaller than that on the copper, and therefore may be neglected.\cite{Yildirim,Sabine} However, even for a free oxygen atom the electron spin orbital coupling turns out to reach of appreciable magnitude: $\xi _{2p}\cong 0.02$ eV,\cite{Herzberg} while for the oxygen O$^{2-}$ ion in oxides one expects the visible enhancement of spin-orbital coupling due to a larger compactness of 2p wave function.\cite{Meier} 
If to account for $\xi _{nl}\propto \langle r^{-3}\rangle _{nl}$ and compare these quantities for the copper ($\langle r^{-3}\rangle _{3d}\approx 6-8$ a.u. \cite{Meier}) and the oxygen ($\langle r^{-3}\rangle _{2p}\approx 4$ a.u.\cite{Walstedt,Meier}) we arrive at a maximum factor two difference in $\xi _{3d}$ and $\xi _{2p}$.
  
 Hereafter we assume a tetragonal symmetry at Cu sites with local coordinate systems as shown in Fig.1. The  global $xyz$ coordinate system is chosen so as Cu$_1$-O-Cu$_2$ plane coincides with $xy$-plane, $x$-axis is directed along Cu$_1$-Cu$_2$ bond (see Fig.1).   In such a case the basic unit vectors ${\bf x},{\bf y},{\bf z}$ can be written in local systems of Cu$_1$ and Cu$_2$ sites as follows: 
 $$
 {\bf x}=(\sin\frac{\theta}{2}, -\cos\frac{\theta}{2}\cos\delta _1,-\cos\frac{\theta}{2}\sin\delta _1); 
 $$
 $$
 {\bf y}=(\cos\frac{\theta}{2}, \sin\frac{\theta}{2}\cos\delta _1,\sin\frac{\theta}{2}\sin\delta _1);
  {\bf z}=(0, \sin\delta _1,\cos\delta _1)
 $$ 
 for Cu$_1$, while for Cu$_2$ site 
$\theta ,\delta _1$ should be replaced by $-\theta ,\delta _2$, respectively.

The exchange integral can be written as follows:
 \begin{equation}
J=\sum_{\{n\},\Gamma}[|c_{\{n\}}(^3\Gamma)|^2\,E_{^3\Gamma}(\{n\})-|c_{\{n\}}(^1\Gamma)|^2\,E_{^1\Gamma}(\{n\})].
\end{equation}

As concerns the DM interaction we deal with two competing contributions. The first is derived as a first order contribution which does not take account of Cu$_{1,2}$ 3d-orbital fluctuations for a ground state 101 configuration. Projecting the spin-orbital coupling (\ref{V_SO}) onto  states (\ref{Psi}) we see that ${\bf \hat\Lambda}^V\cdot {\bf \hat V}$  term is equivalent to purely spin DM coupling with local contributions to Dzyaloshinsky vector
\begin{widetext}
\begin{equation}
D^{(m)}_i=-2i\langle 00|V_{so}(m)|1i\rangle=
-2i\sum_{\{n\}\Gamma_1,\Gamma_2}c_{\{n\}}^*({}^{1}\Gamma_1)c_{\{n\}}({}^{3}\Gamma_2)\langle \Phi_{\{n\};\Gamma _{1} 00} |\Lambda^V_i|\Phi_{\{n\};\Gamma_21i}\rangle \label{DM1}
\end{equation}
\end{widetext}
In all the instances the nonzero contribution to the local Dzyaloshinsky vector is determined solely by the spin-orbital singlet-triplet mixing for the on-site 200, 020, 002 and two-site 110, 011 two-hole configurations, respectively. For on-site two-hole configurations we have ${\bf D}^{(200)}={\bf D}^{(1)}$, ${\bf D}^{(020)}={\bf D}^{(0)}$, ${\bf D}^{(002)}={\bf D}^{(2)}$.

The second contribution,  associated with Cu$_{1,2}$ 3d-orbital fluctuations  within a ground state 101 configuration, is more familiar one and evolves from a second order combined effect of Cu$_{1,2}$ spin-orbital $V_{so}(Cu_{1,2})$ and effective orbitally anisotropic Cu$_1$-Cu$_2$ exchange coupling  
\begin{widetext}
\begin{eqnarray}
D^{(m)}_i=-2i\langle 00|V_{so}(m)|1i\rangle=
-2i\sum_{\Gamma}\frac{\langle \{101\};\Gamma _{s} 00 |\hat\Lambda^V_i|\{101\};\Gamma 1i\rangle \langle \{101\};\Gamma 1i |\hat H_{ex}|\{101\};\Gamma_t1i\rangle}{E_{^3\Gamma _t}(\{101\})-E_{^3\Gamma}(\{101\})} 
 \nonumber \\
 -2i\sum_{\Gamma}\frac{\langle \{101\};\Gamma _{s} 00 |\hat H_{ex}|\{101\};\Gamma 00\rangle \langle \{101\};\Gamma 00 |\hat\Lambda^V_i|\{101\};\Gamma_t1i\rangle}{E_{^1\Gamma _s}(\{101\})-E_{^1\Gamma}(\{101\})}
\end{eqnarray}
\end{widetext}
It should be noted that at variance with the original Moriya approach \cite{Moriya} both spinless and spin-dependent parts of exchange Hamiltonian contribute additively and  comparably to DM coupling, if one takes account of the same magnitude and opposite sign of the singlet-singlet and triplet-triplet exchange matrix elements  on the one hand and  orbital antisymmetry of spin-orbital matrix elements on the other hand.

It is easy to see that the contributions of 002 and 200 configurations to Dzyaloshinsky vector  bear a similarity to 
the respective second type ($\propto V_{so}H_{ex}$) contributions, however, in the former  we deal with spin-orbital coupling for two-hole Cu$_{1,2}$ configurations, while in the latter with that of one-hole  Cu$_{1,2}$ configurations.

\subsection{Copper contribution}
First we address a relatively simple instance of strong rhombic crystal field for intermediate oxygen ion with the crystal field axes oriented along global coordinate $x,y,z$-axes, respectively. It is worth noting that in such a case the oxygen O 2p$_z$ orbital does not play an active role both in symmetric and antisymmetric (DM) exchange interaction as well as Cu 3d$_{yz}$ orbital appears to be inactive in DM interaction due to a zero overlap/transfer with oxygen O 2p orbitals.

For illustration, hereafter we address the first contribution (\ref{DM1}) of two-hole on-site 200, 002 $d^2_{x^2-y^2}$, $d_{x^2-y^2}d_{xy}$, and $d_{x^2-y^2}d_{xz}$ configurations, which do covalently mix with ground state configuration.
Calculating the singlet-triplet mixing matrix elements in the global coordinate system we find all the  components of the local Dzyaloshinsky vectors. The Cu$_1$ contribution turns out to be nonzero only for 200  configuration, and may be written as a sum of several terms. First we present the contribution of singlet $(d^2_{x^2-y^2})^1A_{1g}$ state: 
 \begin{widetext}
$$
D^{(1)}_x=-2i\langle 00|V_{so}(Cu_1)|1x\rangle ={\sqrt{2}}\xi _{3d}\,c_{200}(^1A_{1g})[c_{200}(^3E_{g})\cos\delta _1 -2c_{200}(^3A_{2g})\sin\delta _1 ]\cos\frac{\theta}{2};
$$
$$
D^{(1)}_y=-2i\langle 00|V_{so}(Cu_1)|1y\rangle =-{\sqrt{2}}\xi _{3d}\,c_{200}(^1A_{1g})[c_{200}(^3E_{g})\cos\delta _1 -2c_{200}(^3A_{2g})\sin\delta _1 ]\sin\frac{\theta}{2};
$$
\begin{equation}
 D^{(1)}_z=-2i\langle 00|V_{so}(Cu_1)|1z\rangle =-{\sqrt{2}}\xi _{3d}\,c_{200}(^1A_{1g})[c_{200}(^3E_{g})\sin\delta _1 -2c_{200}(^3A_{2g})\cos\delta _1 ],	
\label{Cu}
\end{equation}
\end{widetext}
where 
$$
c_{200}(^1A_{1g})=-\frac{3}{2\sqrt{2}}t_{pd\sigma}^2\frac{1}{E_{^1A_{1g}}}\left[\frac{\sin^2\frac{\theta}{2}}{\epsilon _x} -\frac{\cos^2\frac{\theta}{2}}{\epsilon _y}\right],
$$ 
$$
c_{200}(^{1,3}A_{2g})=-\frac{\sqrt{3}}{4}t_{pd\sigma}t_{pd\pi}\frac{1}{E_{^{1,3}A_{2g}}}\left(\frac{1}{\epsilon _x} +\frac{1}{\epsilon _y}\right)\sin\theta \cos\delta _1,
$$
$$
c_{200}(^{1,3}E_{g})=-\frac{\sqrt{3}}{4}t_{pd\sigma}t_{pd\pi}\frac{1}{E_{^{1,3}E_{g}}}\left(\frac{1}{\epsilon _x} +\frac{1}{\epsilon _y}\right)\sin\theta \sin\delta _1,
$$
are probability amplitudes for singlet $(d^2_{x^2-y^2})^1A_{1g}$ and singlet/triplet $(d_{x^2-y^2}d_{xy})^{1,3}A_{2g}$, $(d_{x^2-y^2}d_{xz})^{1,3}E_{g}$ 200 configurations in the ground state wave function, respectively.
  $E_{^1A_{1g}}=A+4B+3C$ is the energy of the two-hole copper singlet with $d_{x^2-y^2}^2$ configuration, $E_{^1A_{2g}}=\epsilon _{xy}+A+4B+2C$, $E_{^3A_{2g}}=\epsilon _{xy}+A+4B$, $E_{^1E_{g}}=\epsilon _{xz}+A+B+2C$, $E_{^3E_{g}}=\epsilon _{xz}+A-5B$ are the energies of the two-hole copper terms with $d_{x^2-y^2}d_{xy}$ and $d_{x^2-y^2}d_{xz}$ configurations, respectively, A, B, and C are Racah parameters.  
 Taking into account that $c_{002}(^1A_{1g})=c_{200}(^1A_{1g})$, $c_{002}(^3A_{2g})=c_{200}(^3A_{2g})$, $c_{002}(^3E_{2g})=c_{200}(^3E_{g})$ \cite{remark} we see that the Cu$_2$ contribution to the Dzyaloshinsky vector can be obtained from Exps. (\ref{Cu}), if  $\theta ,\delta _1$  replace by $-\theta ,\delta _2$, respectively. Interestingly to note that
  $D^{(1,2)}_{x,y}\propto \sin\theta \,\sin2\delta _{1,2}$. Both collinear ($\theta = \pi$) and rectangular
 ($\theta = \pi /2$) superexchange geometries appear to be unfavorable for copper contribution to antisymmetric exchange, though in the latter  the result  depends strongly on the relation between the energies of oxygen O 2p$_x$ and O 2p$_y$ orbitals.
 
 Contribution of singlet $(d_{x^2-y^2}d_{xy})^{1}A_{2g}$ and $(d_{x^2-y^2}d_{xz})^{1}E_{g}$ states to the Dzyaloshinsky vector yields  
$$
d^{(1)}_x=d^{(1)}\sin\frac{\theta}{2}\, ,\,\,
d^{(1)}_y=d^{(1)}\cos\frac{\theta}{2}\, , \,\,d^{(1)}_z=0\, ,
$$
where $d^{(1)}=\xi _{3d}(c_{200}(^1A_{2g})c_{200}(^3E_{g})-c_{200}(^1E_{g})c_{200}(^3A_{2g}))$.
Here we deal with a vector directed along Cu$_1$-O bond which   modulus $d^{(1)}\propto \sin^2\theta \sin2\delta _1$ is determined by a partial cancellation of two terms.

It is easy to see that the copper $V_{so}(1)$ contribution to the Dzyaloshinsky vector for two-site 110 and 011 configurations is determined by a $dp$-exchange.
 
\subsection{Oxygen contribution}
In frames of the same assumption regarding the  orientation of rhombic crystal field axes for intermediate oxygen ion the local oxygen contribution to the Dzyaloshinsky vector for one-site 020 configuration  appears to be oriented along local O$_z$ axis and  may be written as follows
\begin{equation}
D^{(0)}_z=-2i\langle 00|V_{so}(O)|1z\rangle ={\sqrt{2}}\xi _{2p}\,c_t(p_xp_y)[c(p_x^2)+c(p_y^2)],	
\label{O}
\end{equation} 
where $c(p^2)=c_{020}(p^2)$ with
$$
c(p_x^2)=-\frac{3}{2\sqrt{2}}t_{pd\sigma}^2\frac{\sin^2\frac{\theta}{2}}{\epsilon _xE_s(p_x^2)};\,
c(p_y^2)=\frac{3}{2\sqrt{2}}t_{pd\sigma}^2\frac{\cos^2\frac{\theta}{2}}{\epsilon _yE_s(p_y^2)};
$$ 
\begin{equation}
c_{t}(p_xp_y)=\frac{3}{8}t_{pd\sigma}^2\left(\frac{1}{\epsilon _x} +\frac{1}{\epsilon _y}\right)\frac{\sin\theta}{E_t(p_xp_y)}
\end{equation}
are probability amplitudes for singlet $p_x^2,p_y^2$ and triplet $p_xp_y$ 020 configurations in the ground state wave function, respectively; $E_s(p_{x,y}^2)=2\epsilon _{x,y}+ F_0+\frac{4}{25}F_2$, $E_t(p_xp_y)=\epsilon _x+\epsilon _y+ F_0-\frac{1}{5}F_2$ are the energies of the oxygen two-hole singlet ($s$) and triplet ($t$) configurations $p_x^2,p_y^2$ and $p_xp_y$, respectively, $F_0$ and $F_2$ are Slater integrals.  
  This vector  can be written as follows \cite{X}
 \begin{equation}
	{\bf D}^{(0)}=D_O(\theta)[{\bf r}_1\times {\bf r}_2],
	\label{x}
\end{equation}
 where ${\bf r}_{1,2}$ are unit radius-vectors along Cu$_1$-O, Cu$_2$-O bonds, respectively, and
 \begin{widetext}
 \begin{equation}
	D_O(\theta)=\frac{9\xi _{2p}t_{pd\sigma}^4}{16}\frac{1}{E_t(p_xp_y)}\left(\frac{1}{\epsilon _x} +\frac{1}{\epsilon _y}\right)\left[\frac{\cos^2\frac{\theta}{2}}{\epsilon _xE_s(p_x^2)}-\frac{\sin^2\frac{\theta}{2}}{\epsilon _yE_s(p_y^2)}\right].\label{OO}
\end{equation}
\end{widetext}
It is worth noting that ${\bf D}^{(0)}$ does not depend on the $\delta _1, \delta _2$ angles. The  $D_O(\theta)$ dependence is expected to be rather  smooth without any singularities for collinear  and rectangular superexchange geometries. 

The local oxygen contribution to the Dzyaloshinsky vector for two-site 110 and 011 configurations  appears to be oriented along local O$_z$ axis as well and  may be written as follows
\begin{widetext}
\begin{equation}
D^{(0)}_z=-2i\langle 00|V_{so}(O)|1z\rangle =\xi \,
([{\bf c}_s(dp)\times {\bf c}_t(dp)]_z+[{\bf c}_s(pd)\times {\bf c}_t(pd)]_z),	
\label{O11}
\end{equation}
\begin{equation}
c_{s,t}(dp_x)=-\frac{\sqrt{3}}{2}\frac{t_{pd\sigma}}{E_{s,t}(dp_x)}\sin\frac{\theta}{2}\,,\,\, 
c_{s,t}(dp_y)=-\frac{\sqrt{3}}{2}\frac{t_{pd\sigma}}{E_{s,t}(dp_y)}\cos\frac{\theta}{2}\,\,,
\end{equation} 
\end{widetext}
where $c_{s,t}(dp)=c_{110}(dp),c_{s,t}(pd)=c_{011}(dp)$ are probability amplitudes for different singlet ($c_s$) and triplet ($c_t$) 110 ($d_{x^2-y^2}p_{x,y}$) and 011 ($p_{x,y}d_{x^2-y^2}$) configurations in the ground state wave function, respectively. The energies $E_{s,t}(dp_{x,y})$  are those for singlet and triplet states of $dp_{x,y}$ configurations, respectively: $E_{s,t}(dp_{x,y})=\epsilon _{x,y}+K_{dpx,y}\pm I_{dpx,y}$, where $K_{dpx,y}$ and $I_{dpx,y}$ are Coulomb and exchange dp-integrals, respectively. It is easy to see that the nonzero contribution to Dzyaloshinsky vector is determined by a direct $dp$-exchange and may be written similarly to (\ref{x}) with
\begin{widetext}
 \begin{equation}
	D_O(\theta)=\frac{3\xi _{2p}t_{pd\sigma}^2}{8}\frac{1}{\epsilon _x\epsilon _y}\left(\frac{I_{dpx}}{\epsilon _x} -\frac{I_{dpy}}{\epsilon _y}\right)\approx \frac{3\xi _{2p}t_{pd\sigma}^2}{8}\frac{1}{\epsilon _x\epsilon _y}\left(\frac{\sin^2\frac{\theta}{2}}{\epsilon _x} -\frac{\cos^2\frac{\theta}{2}}{\epsilon _y}\right)I_{dp\sigma},\label{OO11}
\end{equation}
\end{widetext}
 where we take account only of $dp\sigma$ exchange ($I_{dp\sigma}\propto t^2_{pd\sigma}$).
 
 Thus, the total Dzyaloshinsky vector ${\bf D}$ is a superposition of three contributions ${\bf D}={\bf D}^{(1)}+{\bf D}^{(0)}+{\bf D}^{(2)}$ attached to the respective sites. In general, all the vectors can be oriented differently. Their magnitudes strongly depend on the Cu$_1$-O-Cu$_2$ bond geometry and crystal field effects, however, comparative analysis of Exps. (\ref{Cu}) and (\ref{OO}) given estimations for different parameters typical for cuprates\cite{Eskes} ($t_{pd\sigma}\approx 1.5$ eV, $t_{pd\pi}\approx 0.7$ eV, $A=6.5$ eV, $B=0.15$ eV, $C=0.58$ eV, $F_0=5$ eV, $F_2=6$ eV) evidences that copper and oxygen Dzyaloshinsky vectors can be of comparable magnitude.

\subsection{Dzyaloshinsky-Moriya coupling in La$_2$CuO$_4$}
 
 The DM coupling and magnetic anisotropy in La$_2$CuO$_4$ and related compounds
has attracted considerable attention in 90-ths (see, e.g., Refs.\onlinecite{Coffey,Koshibae,Shekhtman}), and is still debated in the literature.\cite{Tsukada,Kataev} 
In the low-temperature tetragonal (LTT) and orthorhombic (LTO) phases of La$_2$CuO$_4$, the
oxygen octahedra surrounding each copper ion rotate by a
small tilting angle ($\delta _{LTT}\approx 3^0,\delta _{LTO}\approx 5^0$) relative to their location in the high-temperature tetragonal phase.
The structural distortion  allows for the appearance of the antisymmetric Dzyaloshinsky-Moriya interaction.
In terms of our choice for structural parameters to describe the Cu$_1$-O-Cu$_2$ bond we have for LTT phase:
$ \theta =\pi; \delta _1=\delta _2=\frac{\pi}{2}\pm \delta _{LTT}$ for bonds oriented perpendicular to the tilting plane, and $ \theta =\pm(\pi -2\delta _{LTT}); \delta _1=\delta _2=\frac{\pi}{2}$ for bonds oriented parallel to the tilting plane. It means that all the local Dzyaloshinsky vectors turn into zero for the former bonds, and turn out to be perpendicular to the tilting plane for the latter bonds. For LTO phase:$ \theta =\pm(\pi -\sqrt{2}\delta _{LTO}); \delta _1=\delta _2=\frac{\pi}{2}\pm \delta _{LTO}$. In such a case the largest ($\propto \delta _{LTO}$) component of the local Dzyaloshinsky vectors (z-component in our notation) turn out to be oriented perpendicular to the Cu$_1$-O-Cu$_2$ bond plane. Other two components of the local Dzyaloshinsky vectors are fairly small: that of perpendicular to CuO$_2$ plane (y-component in our notation) is of the order of $\delta _{LTO}^2$, while that of oriented along Cu$_1$-Cu$_2$ bond
 axis (x-components in our notation) is of the order of $\delta _{LTO}^3$.

 \section{DM coupled Cu$_1$-O-Cu$_2$ bond in external fields}
 \subsection{Uniform external magnetic field}
 Application of an uniform external magnetic field ${\bf h}_S$ will produce a staggered spin polarization in the antiferromagnetically coupled Cu$_1$-Cu$_2$ pair
\begin{equation}
	\langle {\bf V}_{12}\rangle={\bf L}=-\frac{1}{J_{12}^2}[\sum_i{\bf D}_{12}^{(i)}\times {\bf h}^S]={\bf \stackrel{\leftrightarrow}{\chi}}^{VS}{\bf h}^S
\label{V}
\end{equation}
with antisymmetric $VS$-susceptibility tensor: $\chi _{\alpha\beta}^{VS}=-\chi _{\beta\alpha}^{VS}$. One sees that the sense of a staggered spin polarization, or antiferromagnetic vector, depends on that of  Dzyaloshinsky vector.\cite{Moskvin2}  The $VS$ coupling
results in many interesting effects for the DM systems, in particular, the "field-induced gap" phenomena in 1D s=1/2 antiferromagnetic Heisenberg system with alternating DM coupling.\cite{Affleck} Approximately, the phenomenon is described by a so called $staggered$  s=1/2 antiferromagnetic Heisenberg model with the Hamiltonian
\begin{equation}
\hat H=J\sum_i (\hat{\bf s}_i\cdot \hat{\bf s}_{i+1})-h_u\hat s_{iz}-(-1)^ih_s\hat s_{ix} \, ,\label{staggered}
\end{equation}  
which includes the effective uniform field $h_u$ and the induced staggered field $h_s\propto h_u$ perpendicular both to the applied uniform magnetic field and Dzyaloshinsky vector. 


\subsection{Staggered external field}

 Application of a staggered field ${\bf h}^V$ for an antiferromagnetically coupled Cu$_1$-Cu$_2$ pair will produce a local spin polarization both on copper and oxygen sites 
\begin{equation}
	\langle {\bf S}_{i}\rangle=\frac{1}{J_{12}^2}[{\bf D}_{12}^{(i)}\times {\bf h}^V]={\bf \stackrel{\leftrightarrow}{\chi}}^{SV}(i){\bf h}^V,
\label{S}
\end{equation}
 which can be detected by different site-sensitive methods including neutron diffraction and, mainly, by nuclear magnetic resonance. It should be noted that $SV$-susceptibility tensor is the antisymmetric one: $\chi _{\alpha\beta}^{SV}=-\chi _{\beta\alpha}^{SV}$. 
 Strictly speaking, both formulas (\ref{V}),(\ref{S}) work well only in a paramagnetic regime and for relatively weak external fields. 
 
 Above we addressed a single Cu$_1$-O-Cu$_2$ bond, where, despite a site location, the direction and magnitude of Dzyaloshinsky vector  depends strongly on the bond strength and geometry. 
 It is clear that a site rather than a bond location of DM vectors would result in a revisit of conventional symmetry considerations and of magnetic structure in weak ferro- and antiferromagnets. Interestingly that the expression (\ref{S}) predicts the effects of a constructive or destructive (frustration) interference of copper spin polarizations in 1D, 2D, and 3D lattices depending on the relative sign of Dzyaloshinsky vectors and staggered fields for nearest neighbours. It should be noted that with the destructive interference the local copper spin polarization may turn into zero and DM coupling will manifest itself only through the oxygen spin polarization.  
\begin{figure}[h]
\includegraphics[width=8.5cm,angle=0]{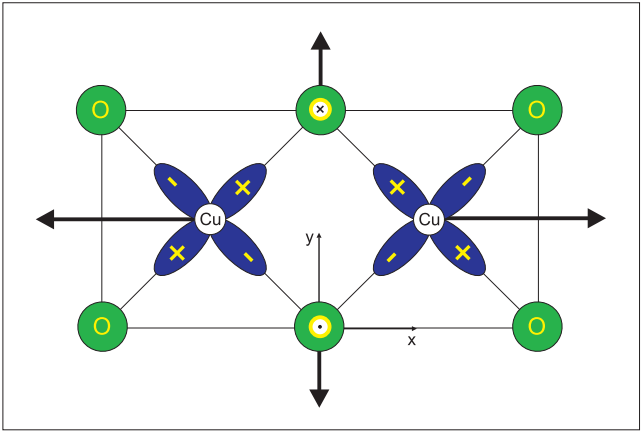}
\caption{The fragment of a typical edge-shared CuO$_2$ chain with copper and oxygen spin orientation under staggered field applied along x-direction. Note the antiparallel orientation of oxygen Dzyaloshinsky vectors.}
\label{fig2}
\end{figure} 
 Another interesting manifestation of oxygen Dzyaloshinsky-Moriya antisymmetric exchange coupling concerns the edge-shared CuO$_2$ chains (see Fig.2), ubiquitous in many cuprates, where we deal with an exact compensation of copper contributions to Dzyaloshinsky vectors and the possibility to observe the effects of uncompensated though oppositely directed  local oxygen contributions.  Applying the staggered field along chain direction (O$_x$) we arrive in accordance with Exp.(\ref{S}) at a $weakly$ antiferromagnetic O$_y$ ordering of oxygen spin polarizations which magnitude is expected to be strongly enhanced due to usually small magnitudes of 90$^o$ symmetric superexchange. It should be emphasized that the summary  Dzyaloshinsky vector turns into zero  hence in terms of a conventional approach to DM theory we miss the anomalous oxygen spin polarization effect. In this connection it is worth noting the neutron diffraction data by Chung {\it et al.}\cite{Chung}  which unambiguously evidence the oxygen momentum formation and canting in  edge shared CuO$_2$ chain cuprate Li$_2$CuO$_2$.

\subsection{Intermediate oxygen $^{17}$O Knight shift as an effective tool  to inspect DM coupling for Cu-O-Cu bonds}
As an important by-product of the cuprate activity we arrived at a significant progress in different experimental methods and new opportunities to elucidate subtle details of electron and spin structure. In particular, it concerns the oxygen $^{17}$O NMR-NQR as an unique local probe to study the charge and spin densities on oxygen sites.\cite{Walstedt}
 In this connection we point to papers by R. Walstedt {\it et al}.\cite{Walstedt} as a first direct observation of  anomalous oxygen hyperfine interactions in  generic cuprate La$_2$CuO$_4$. With the approaching  transition to the ordered magnetic phase,  the authors observed anomalously large negative $^{17}$O Knight shift for planar oxygens which anisotropy resembled that of weak ferromagnetism in this cuprate. The giant shift was observed only when external field was parallel to the local Cu-O-Cu bond axis (PL1 lines) or perpendicular to CuO$_2$ plane. The effect was not observed  for PL2 lines which correspond to oxygens in the local Cu-O-Cu bonds which axis is perpendicular to in-plane external field.  The data were interpreted as an indication of oxygen spin polarization due to a local Dzyaloshinsky-Moriya antisymmetric exchange coupling. However, either interpretation of NMR-NQR data in such low-symmetry systems as La$_2$CuO$_4$ needs in a thorough analysis of transferred hyperfine interactions and a revisit of some textbook results  being typical for the model high-symmetry systems. First we draw attention to spin-dipole hyperfine interactions for O 2p-holes which are main participants of Cu$_1$-O-Cu$_2$ bonding. 
 Starting from a conventional formula for a spin-dipole contribution to local field
 $$
	{\bf \cal H}_n =-g_s\mu_B\sum_i\frac{3({\bf r}_i\cdot{\bf s}_i){\bf r}_i-r_i^2{\bf s}_i}{r_i^5}
$$
and making use of an expression for  appropriate  matrix element 
$$
	\langle p_i| \frac{3x_{\alpha}x_{\beta}-r^2\delta_{\alpha\beta}}{r^5}|p_j\rangle =-\frac{2}{5}\left\langle \frac{1}{r^3}\right\rangle _{2p}\langle p_i| 3\widetilde{l_{\alpha}l_{\beta}}-2\delta_{\alpha\beta}|p_j\rangle =
$$
\begin{equation}	
	\frac{2}{5}\left\langle \frac{1}{r^3}\right\rangle _{2p}(\frac{3}{2}\delta_{\alpha i}\delta_{\beta j}+\frac{3}{2}\delta_{\alpha j}\delta_{\beta i}-\delta_{\alpha\beta}\delta_{ij})
\label{pipj}
\end{equation}  
we present a local field on the $^{17}$O nucleus in Cu$_1$-O-Cu$_2$ system as a sum of ferro- and antiferromagnetic contributions as follows\cite{Moskvin2} 
 \begin{equation}
{\bf \cal H}_n={\bf \stackrel{\leftrightarrow}{A}}^S\cdot \langle{\bf \hat S}\rangle+{\bf \stackrel{\leftrightarrow}{A}}^V\cdot \langle{\bf \hat V}\rangle
\end{equation}
where 
$$
{\bf \stackrel{\leftrightarrow}{A}}^S={\bf \stackrel{\leftrightarrow}{A}}^S(dp)+{\bf \stackrel{\leftrightarrow}{A}}^S(pd);\, {\bf \stackrel{\leftrightarrow}{A}}^V={\bf \stackrel{\leftrightarrow}{A}}^V(pd)-{\bf \stackrel{\leftrightarrow}{A}}^V(dp)\, ,
$$

$$
A^S_{ij}(dp)=A_p^{(0)}[3c_t(dp_i)c_t(dp_j)-|{\bf c}_t(dp)|^2\delta_{ij}]\, ,
$$
$$
A^S_{ij}(pd)=A_p^{(0)}[3c_t(p_id)c_t(p_jd)-|{\bf c}_t(pd)|^2\delta_{ij}]\, ,
$$ 
$$
A^V_{ij}(dp)=A_p^{(0)}[3\widetilde{c_s(dp_i)c_t(dp_j)}-({\bf c}_s(dp)\cdot{\bf c}_t(dp))\delta_{ij}]\, ,
$$
$$
A^V_{ij}(pd)=A_p^{(0)}[3\widetilde{c_s(p_id)c_t(p_jd)}-({\bf c}_s(pd)\cdot{\bf c}_t(pd))\delta_{ij}]\, ,
$$
 where $A_{p}^{(0)}=\frac{2}{5}g_s\mu_B\left\langle \frac{1}{r^3}\right\rangle _{2p}$, the tilde points to a symmetrization. Thus, along with a conventional textbook  ferromagnetic ($\propto \langle{\bf \hat S}\rangle$) transferred hyperfine contribution to local field which simply mirrors a sum total of two Cu-O bonds,  we arrive at an additional unconventional antiferromagnetic difference ($\propto \langle{\bf \hat V}\rangle$) contribution which symmetry and  magnitude strongly depend on the orientation of the oxygen crystal field axes and  Cu$_1$-O-Cu$_2$ bonding angle. With the Cu$_1$-O-Cu$_2$ geometry shown in Fig.1 we arrive at a diagonal ${\bf \stackrel{\leftrightarrow}{A}}^S$ tensor: 
$$
A^S_{xx}=2A_{p}(3\sin^2\frac{\theta}{2} -1);\,A^S_{yy}=2A_{p}(3\cos^2\frac{\theta}{2} -1);\,A^S_{zz}=-2A_{p},
$$
and the only nonzero components of ${\bf \stackrel{\leftrightarrow}{A}}^V$ tensor: 
$$
A^V_{xy}=A^V_{yx}=3A_{p}\sin\theta 
$$
 with
$$
A_{p}=\frac{3}{4}\left(\frac{t_{dp\sigma}}{\epsilon_{p}}\right)^2A_{p}^{0}=f_{\sigma}A_{p}^{0},
$$
where $f_{\sigma}$ is the parameter of a transferred spin density and we made use of a simple approximation $E_{s,t}(dp_{x,y})\approx \epsilon_{p}$. 
Generally speaking, we should take into account an additional contribution of magneto-dipole hyperfine interactions.

The two-term structure of oxygen local field implies a two-term  S-V structure of the $^{17}$O Knight shift
$$
^{17}K={\bf \stackrel{\leftrightarrow}{A}}^S{\bf \stackrel{\leftrightarrow}{\chi}}^{SS} +
{\bf \stackrel{\leftrightarrow}{A}}^V{\bf \stackrel{\leftrightarrow}{\chi}}^{VS}
$$
that points to Knight shift as an effective tool to inspect both uniform and staggered spin polarization. 
 The existence of antiferromagnetic term in oxygen hyperfine interactions yields a rather simple explanation of the $^{17}$O Knight shift anomalies in La$_2$CuO$_4$\cite{Walstedt} as a result of the external field induced  staggered spin polarization
$\langle{\bf \hat V}\rangle =\bf L = {\bf \stackrel{\leftrightarrow}{\chi}}^{VS} {\bf \cal H}_{ext}$. Indeed, "our" local $y$ axis for Cu$_1$-O-Cu$_2$ bond corresponds to  the crystal tetragonal $c$-axis oriented perpendicular to CuO$_2$ planes both in LTO and LTT phases of La$_2$CuO$_4$ while $x$-axis does to local Cu-O-Cu bond axis. It means that for the geometry of the experiment by Walstedt {\it et al.}\cite{Walstedt} (the crystal is oriented so that the external uniform field is either $\parallel$ or $\perp$ to the local Cu-O-Cu bond axis) the antiferromagnetic contribution to $^{17}$O Knight shift will be observed  only a) for oxygens in Cu$_1$-O-Cu$_2$ bonds oriented  along external field or b) for external field along tetragonal $c$-axis. Experimental data\cite{Walstedt} agree with staggered magnetization along the tetragonal $c$-axis in the former and along the rhombic $c$-axis (tetragonal $ab$-axis) in the latter.
Interestingly, the sizeable effect  has been observed in  La$_2$CuO$_4$ for temperatures $T\sim 500$ K that is essentially higher than $T_N \approx 300$ K.  Given $L=1$, $A_{p}^{(0)}\approx 100$ kG/spin,\cite{Walstedt} $|\sin\theta | \approx 0.1$, and $f_{\sigma}\approx 20\%$ we obtain $\approx 6$ kG as a maximal value of a low-temperature antiferromagnetic contribution to hyperfine field which is parallel to external magnetic field. This value agrees with a low-temperature extrapolation of high-temperature experimental data by Walstedt {\it et al.}\cite{Walstedt} Similar effect of anomalous $^{13}$C Knight shift has recently been observed in copper pyrimidine dinitrate  [CuPM(NO$_3$)$_2$(H$_2$O)$_2$]$_n$, a one-dimensional S=1/2 antiferromagnet with alternating local symmetry.\cite{CuPM}
However, the authors did take into account  only the inter-site  magneto-dipole contribution to ${\bf \stackrel{\leftrightarrow}{A}}^V$ tensor that  questions their quantitative conclusions regarding the "giant" spin canting.

The ferro-antiferromagnetic S-V structure  of local field on the nucleus of an intermediate oxygen ion in a  Cu$_1$-O-Cu$_2$ triad points to $^{17}$O NMR as, probably, the only experimental technique to measure both the value, direction, and the sense of Dzyaloshinsky vector. The latter possibility was realized earlier with $^{19}$F NMR for  weak ferromagnet FeF$_3$.\cite{Moskvin2} 
The experimental observation of antiferromagnetic contribution to the Knight shift provides probably the sole way to find out the problem of the sense of Dzyaloshinsky vector in cuprates. For instance, the negative sign of $^{17}$O Knight shift in La$_2$CuO$_4$\cite{Walstedt}  points to a negative sign of ${\bf \stackrel{\leftrightarrow}{\chi}}^{VS}$ for Cu$_1$-O-Cu$_2$ triad with $A_{xy}^V>0$, hence to a positive sense of $z$-component of  the summary Dzyaloshinsky vector in Cu$_1$-O-Cu$_2$ triad with geometry shown in Fig.1 given $\theta \leq \pi$, $\delta_1=\delta_2\approx \pi /2$.
This agrees with theoretical predictions both for copper and oxygen contributions based on Exps.(\ref{Cu}) and (\ref{OO}). It should be emphasized that the above effect is determined by the summary  Dzyaloshinsky vector in Cu$_1$-O-Cu$_2$ triad rather than by a local oxygen "weak-ferromagnetic" polarization as it was proposed by Walstedt {\it et.al.} \cite{Walstedt}

\section{Symmetric spin anisotropy}
Symmetric spin anisotropy for a Cu$_1$-Cu$_2$ pair is described by an effective Hamiltonian
\begin{equation}
\hat H_{an}=\frac{1}{4}\hat{\bf S}{\bf \stackrel{\leftrightarrow}{K}}^S \hat{\bf S}-\frac{1}{4}\hat{\bf V}{\bf \stackrel{\leftrightarrow}{K}}^V \hat{\bf V}
\label{Han}
\end{equation}
with kinematic relations (\ref{SV}).

Depending on the sign of the anisotropy constants we arrive at two types of spin configurations minimizing the energy of spin anisotropy: the conventional twofold degenerated ferromagnetic state or unconventional multiple degenerated  antiferromagnetic state. As a relevant illustrative example one might refer to the axial anisotropy: $\hat H_{an}=K\hat S_z^2$ which stabilizes the $|1\pm 1\rangle $ doublet given $K<0$ or a set of superposition states 
\begin{equation}
\Psi _{\alpha ,\phi} =\cos\alpha |00\rangle +e^{i\phi}\sin\alpha |10\rangle	
\label{Psi1}
\end{equation}
 given $K>0$. The latter  incorporates the limiting configurations $|00\rangle$ and $|10\rangle$, the N\'{e}el  ($\frac{1}{\sqrt{2}}(|00\rangle \pm|10\rangle$) and DM doublets ($\frac{1}{\sqrt{2}}(|00\rangle \pm i|10\rangle$), and any other their superpositions.

As usual, the term is processed in frames of a number of simple model approximations. First, instead of the generalized form (\ref{Han}) one addresses a pseudodipole anisotropy 
\begin{equation}
\hat H_{an}=\hat{\bf s}_1{\bf \stackrel{\leftrightarrow}{K}}_{12} \,\hat{\bf s}_2 \,.
\label{Han1}
\end{equation} 
Second, one makes use of a trivial local MFA approach which applicability for s=1/2 spin systems is questionable. Indeed, calculating the classical energy of the pseudodipole two-ion anisotropy $K\langle \hat s_{1z}\rangle \langle \hat s_{2z}\rangle = \frac{1}{4}K(\langle \hat S_{z}\rangle ^2- \langle \hat V_{z}\rangle ^2 )$ and respective quantum energy
$K\langle \hat s_{1z} \hat s_{2z}\rangle =  \frac{1}{4}K(\langle \hat S_{z}^2\rangle - \langle \hat V_{z}^2\rangle )$ for a two-ion state $\cos\alpha |00\rangle +\sin\alpha |1n\rangle$ induced by a N\'{e}el-like staggered field ${\bf h}^V\parallel {\bf n}$ we arrive at
$E_{class}=- \frac{1}{4}K\sin2\alpha \cdot n_z^2$ and $E_{quant}= \frac{1}{4}K(1-2\sin^2\alpha  \cdot n_z^2)$, respectively, that  evidences the crucial importance of quantum effects when addressing the numerical aspect of spin anisotropy. Note that the mean value $\langle \hat V_{z}^2\rangle $ reaches the maximum (=1) on a set of superposition states (\ref{Psi1}), while $\langle \hat V_{z}\rangle ^2$ does on a single N\'{e}el state $\Psi _{\alpha =\pi /4,\phi =0}$.

\subsection{Effective symmetric spin anisotropy due to DM interaction}
When one says about an effective  spin anisotropy due to DM interaction one usually addresses a simple classical two-sublattice weak ferromagnet where the free energy has a minimum when both ferro- ($\propto \langle \hat {\bf S}\rangle $) and antiferromagnetic ($\propto \langle \hat {\bf V}\rangle $) vectors, being perpendicular to each other, lie in the plane perpendicular to Dzyaloshinsky vector ${\bf D}$. However, the issue is rather involved and appeared for a long time to be hotly debated.\cite{Shekhtman,debate,Zheludev,vanadate} In our opinion, first of all we should define what spin anisotropy is. Indeed, the description of any spin system implies  the free energy $\Phi$ depends on a set of vectorial order parameters (e.g.,$\langle \hat {\bf S}\rangle ,\propto \langle \hat {\bf V}\rangle ,\propto \langle \hat {\bf T}\rangle $) under  kinematic constraint, rather than a single magnetic moment as in a simple ferromagnet, that can make the orientational dependence of $\Phi$ extremely involved. Such a situation needs in a careful analysis of respective spin Hamiltonian with a choice of proper approximations.

Effective symmetric spin anisotropy due to DM interaction can be easily derived as a second order perturbation correction due to DM coupling as follows:
$$
\hat H_{an}^{DM}=\hat P {\hat H}_{DM}{\hat R}{\hat H}_{DM}\hat P \, ,
$$
where $\hat P $ is the projection operator projecting on the ground manifold and ${\hat R}=(1-{\hat P})/(E_0-{\hat H}_0)$. For antiferromagnetically coupled spin 1/2 pair $\hat H_{an}^{DM}$ may be written as follows:
$$
\hat H_{an}^{DM}=\sum_{ij}\Delta K^V_{ij} {\hat V}_i{\hat V}_j
$$
with
$\Delta K^V_{ij} =\frac{1}{8J} D_{i}D_{j}$ provided $|{\bf D}|\ll J$. We see that in frames of a simple MFA approach this anisotropy stabilizes a N\'{e}el state with $\langle \hat {\bf V}\rangle \perp {\bf D}$. However, in fact we deal with an MFA artefact. Indeed, let examine the second order perturbation correction to the ground state energy of an antiferromagnetically coupled spin 1/2 pair in  a N\'{e}el-like staggered field ${\bf h}^V\parallel {\bf n}$: 
\begin{equation}
E_{an}^{DM}=-\frac{|{\bf D}\cdot {\bf n}|^2}{4(E_{\parallel}-E_g)}-\frac{|{\bf D}\times {\bf n}|^2}{4(E_{\perp}-E_g)}\cos^2\alpha \, ,
\label{an}
\end{equation}
where $E_{\perp}=J; E_{\parallel}=J\,\cos^2\alpha +h^V\,\sin2\alpha; E_{g}=J\,\sin^2\alpha -h^V\,\sin2\alpha$. First term in (\ref{an}) stabilizes ${\bf n}\parallel{\bf D}$ configuration while the second one does the ${\bf n}\perp{\bf D}$
configuration. Interestingly that $(E_{\parallel}-E_g)\cos^2\alpha = (E_{\perp}-E_g)$, that is for any  staggered field $E_{an}^{DM}$  does not  depend on its orientation. In other words, at variance with a simple MFA approach, the DM contribution to the energy of anisotropy for an exchange coupled spin 1/2 pair in a staggered field turns into zero. The conclusion proves to be correct in the limit of a zero field as well.

\subsection{Microscopics of symmetric spin anisotropy}
Anyway, the $\hat H_{an}^{DM}$ term has not to be included into an effective spin anisotropy Hamiltonian (\ref{HS}).
As concerns the true symmetric two-ion spin anisotropy (pseudodipole, or exchange anisotropy), its magnitude can be obtained if we take into account all other effects quadratic in spin-orbital coupling. 


At variance with the effective DM spin-Hamiltonian the symmetric spin anisotropy evolves from the higher-order perturbation effects that makes its analysis even more involved and leads to many misleading estimations. Similarly to DM interaction we deal with two competing contributions. The first is derived as a lowest order contribution which does not take account of orbital fluctuations for Cu$_{1,2}$ 3d states.  
 To this end we take into account the effects of spin-orbital mixing for ground state  singlet and triplet 101 configurations perturbed by covalent effects. Assuming the validity of conventional perturbation  series we arrive at a modified expression for respective functions as follows
\begin{widetext}
\begin{equation}
\Psi_{101;SM}=\Phi_{101;SM}+\sum_{\{n\}\Gamma}c_{\{n\}}({}^{2S+1}\Gamma)\left[\Phi_{\{n\};\Gamma SM}-\sum_{S^{\prime}M^{\prime}\Gamma^{\prime}}\frac{\langle \{n\};\Gamma^{\prime} S^{\prime}M^{\prime}|V_{so}|\{n\};\Gamma SM\rangle}{E_{^{2S^{\prime}+1}\Gamma^{\prime}}(\{n\})-E_{^{2S+1}\Gamma_0}(101)}\Phi_{\{n\};\Gamma^{\prime} S^{\prime}M^{\prime}}\right]\label{WF}
\end{equation}
\end{widetext}
and arrive at the expressions for the tensorial anisotropy  parameters  as follows:
\begin{widetext}
 \begin{equation}
K_{ij}^S=\sum_{\{n\}\Gamma_1,\Gamma_2,\Gamma^{\prime}}c_{\{n\}}^*({}^{3}\Gamma_1)c_{\{n\}}({}^{3}\Gamma_2)\frac{\langle \{n\};^3\Gamma_1 |\Lambda^S_i|\{n\};^3\Gamma^{\prime}\rangle \langle \{n\};^3\Gamma^{\prime} |\Lambda^S_j|\{n\};^3\Gamma_2\rangle}{E_{^{3}\Gamma^{\prime}}(\{n\})-E_{^{3}\Gamma_0}(101)},\label{K^S}
\end{equation}
\begin{equation}
K_{ij}^V=\sum_{\{n\}\Gamma_1,\Gamma_2,\Gamma^{\prime}}c_{\{n\}}^*({}^{3}\Gamma_1)c_{\{n\}}({}^{3}\Gamma_2)\frac{\langle \{n\};^3\Gamma_1 |\Lambda^V_i|\{n\};^1\Gamma^{\prime}\rangle \langle \{n\};^1\Gamma^{\prime} |\Lambda^V_j|\{n\};^3\Gamma_2\rangle}{E_{^{1}\Gamma^{\prime}}(\{n\})-E_{^{3}\Gamma_0}(101)}.\label{K^V}
\end{equation}
\end{widetext}
 We see that $K_{ij}^S$ and $K_{ij}^V$ are determined by the triplet-triplet and singlet-triplet mixing, respectively. Interestingly that for nonzero orbital matrix elements in (\ref{K^S}) and (\ref{K^V}) we find
$$
\langle \{n\};^3\Gamma_1 |\Lambda^S_i|\{n\};^3\Gamma^{\prime}\rangle =\langle \{n\};^3\Gamma_1 |\Lambda^V_i|\{n\};^1\Gamma^{\prime}\rangle 
$$ 
hence $K_{ij}^S=K_{ij}^V$, if we will  suppose that $E_{^{1}\Gamma^{\prime}}(\{n\})=E_{^{3}\Gamma^{\prime}}(\{n\})$, that is neglect the singlet-triplet splitting for $\Gamma^{\prime}$ terms, or respective exchange effects.   
It should be noted that the contribution of two-hole on-site 200, 002, and 020 configurations in (\ref{K^S}) and (\ref{K^V}) is actually related to an on-site, or single ion spin anisotropy. 
Thus we conclude that, strictly speaking, a simple two-site pseudodipole form of symmetric anisotropy (\ref{1}) fails to capture correctly all the features of spin anisotropy in our three-center two-hole system. Firstly it concerns the quantitative predictions and estimations.
The contribution of two-hole two-site 110 and 011 configurations to spin anisotropy turns out to be nonzero, only if pd-exchange is taken into account.
 

The second contribution is associated with orbital fluctuations for Cu$_{1,2}$ 3d states within a ground state 101 configuration, and evolves from a third order combined effect of Cu$_{1,2}$ spin-orbital $V_{so}(Cu_{1,2})$ and effective Cu$_1$-Cu$_2$ exchange coupling (see, e.g., a detailed analysis in Ref.\onlinecite{Moskvin1}).  
Thus, we see that any decisive conclusions regarding the quantitative estimations of symmetric spin anisotropy imply a thorough analysis of numerous competing contributions.

\section{Conclusions}
    
 In conclusion, we have revisited a problem of  Dzyaloshinsky-Moriya antisymmetric exchange coupling in cuprates specifying the local spin-orbital contributions to Dzyaloshinsky vector focusing on the oxygen term. We have applied a scheme that provides an optimal way to account for intra-atomic electron correlations and separate the local contributions to Dzyaloshinsky vector. 
 The Dzyaloshinsky vector and respective weak ferromagnetic momentum is shown to be a superposition of comparable and, sometimes, competing local Cu and O contributions.  
  The intermediate oxygen $^{17}$O Knight shift is shown to be an effective tool to inspect the effects of Dzyaloshinsky-Moriya coupling in an external magnetic field. Anisotropic antiferromagnetic contribution to $^{17}$K explains the anomalies observed in La$_2$CuO$_4$.\cite{Walstedt} We predict the effect of $strong$ oxygen weak antiferromagnetism in edge-shared CuO$_2$ chains due to uncompensated oxygen Dzyaloshinsky vectors. Its experimental observation  could provide a direct evidence of the oxygen DM coupling. We revisit the effects of symmetric spin anisotropy, in particular, those directly induced by Dzyaloshinsky-Moriya coupling.

 Finally it should be noted that the anionic contribution to Dzyaloshinsky vector is crucial for the very existence 
of the DM coupling in the pair of rare-earth ions (see, e.g., Yb$^{3+}$-As$^{4-}$-Yb$^{3+}$ triads in Yb$_4$As$_3$\cite{YbAs}) because very strong spin-orbital coupling for rare-earth ions is diagonalized within a ground state multiplet.
  
\begin{acknowledgments}
 
 I thank R. Walstedt for stimulating and encouraging discussion, H. Eschrig, M. Richter, and S.-L. Drechsler for their interest and useful discussion. I would like to thank Leibniz-Institut f\"ur Festk\"orper- und Werkstoffforschung Dresden where part of this work was made for  hospitality. Partial support by  CRDF Grant No. REC-005, RFBR grants Nos. 04-02-96077, 06-02-17242, and 06-03-90893 is  acknowledged.
\end{acknowledgments}


\begin{thebibliography}{99}
\bibitem{remark}
This is true to within a replacement $\delta _1 \leftrightarrow \delta _2$. It should be noted that the probability amplitudes for triplet 200 and 002 configurations are of the same sign due to double-minus effect: 1) $\theta \leftrightarrow -\theta $ replacement, and 2)the antisymmetry of orbital functions: 200-function $\propto d_{x^2-y^2}(1)d_{xy}(2)$, while 002-function $\propto d_{x^2-y^2}(2)d_{xy}(1)$.
\end{thebibliography}
\end{document}